\documentclass[notitlepage,prb,
twocolumn,
longbibliography,superscriptaddress,amsmath,amssymb]{revtex4-2}

\usepackage{graphicx}  % needed for figures
\usepackage{dcolumn}   % needed for some tables
\usepackage{bm}        % for math
\usepackage{verbatim}   % for math
\usepackage{bbold}
\usepackage{color}
\usepackage{xcolor}
\usepackage{amsthm}
\usepackage{amsmath}
\usepackage{ulem}
\usepackage{eufrak}

\usepackage{amsmath}
\usepackage{amsthm, amssymb}
\usepackage{slashed} 
\usepackage{graphicx}
\usepackage{xcolor}
\usepackage{bm}
\usepackage[colorlinks=true,citecolor=blue,linkcolor=magenta]{hyperref}
\usepackage{dsfont}
\usepackage{changepage}
\usepackage{array}

\def\be{\begin{equation}}
\def\ee{\end{equation}}
\def\ba{\begin{align}}
\def\ea{\end{align}}
\def\>{\rangle}
\def\<{\langle}

\def\bea{\begin{eqnarray}}
\def\eea{\end{eqnarray}}
\def \T*{\text{T}^*}

\def \K3C60{\text{K}_3\text{C}_{60}}
\def \ETBr{\kappa\text{-(ET})_2\text{Cu}[\text{N(CN)}_2]\text{Br}}

\def\>{\rangle}
\def\<{\langle}

\theoremstyle{definition}

\theoremstyle{remark}

\begin{document}

%\mz{Maybe period-doubled superconductivity is better pitch for title}
\title{Photo-induced Superconductivity =    Discrete Time Crystal?}
%\title{Photo-induced Superconductivity: a  Discrete Time Crystal?}
%\title{Photo-induced Superconductivity =    Discrete Time Crystal?}

\author{Zhehao Dai}
\affiliation{
University of California, Berkeley, CA 94720, USA}

\author{Vibhu Ravindran}
\affiliation{
University of California, Berkeley, CA 94720, USA}

\author{Norman Y. Yao}
\affiliation{
University of California, Berkeley, CA 94720, USA}

\author{Michael P. Zaletel}
\affiliation{
University of California, Berkeley, CA 94720, USA}

\date{\today}

\begin{abstract}
We propose that periodic driving can stabilize a new type of order, ``period-doubled superconductivity'',  in which a superconducting order parameter oscillates at half the frequency of the drive.
Despite having a zero time-averaged order parameter, the ordered state exhibits perfect conductivity and a Meissner effect. Our theory predicts that this phase may be realized as the steady state of materials shown to exhibit photo-induced superconductivity. We propose to detect the period-doubled oscillation of the order parameter by utilizing a Josephson junction between a photo-induced superconductor and a conventional superconductor. Our theory can also be realized via parametric driving of cold bosonic atoms in an optical lattice.
\end{abstract}

\maketitle

% floquet engineering power-ful tool explored in many expts
% simplest and prominent is DTC
% another example where driving does cool stuff is photo-indcued SC
% here reveal a possible connection between the two
% analyze or understand photo induced sc within the framework or as a activated DTC.
% if $U(1)$ is exactly preserved could be tc for ever
% every floquet 
%\section{Introduction}

% TODO: norm is going to do once over and make small changes and ask questions
% MIKE needs to read dissipation and noise section
% MIKE needs to add in discussion about relationship to Time crystal
% Zhehao needs to update figure （done）
% Zhehao needs to understand whether the Shapiro is already PDSC

%\mz{I think we should edit this paragraph to better anticipate Patrick's gut reaction: isn't ``time-crystal'' just a hype-filled rebranding of parametric resonance?}

%\zd{superconductivity in high-temperature driven nonequilibrium systems? Cavalleri} 

\section{Introduction}
Superconductivity is associated with a coherent supercurrent, corresponding to the flow of a Cooper-pair condensate.
While conventionally found at low temperatures in thermal equilibrium, recent experiments have observed so-called ``photo-induced superconductivity'' in a number of strongly correlated materials out of equilibrium, including cuprate high-temperature superconductors~\cite{PhysRevB.89.184516,hu2014optically,cavalleri2018photo,PhysRevX.10.011053}, organic molecular materials~\cite{PhysRevX.10.031028,PhysRevLett.127.197002}, and a fulleride~\cite{mitrano2016possible}.
After an intense laser pulse, the materials  exhibit a transient superconducting-like response in their optical conductivity at temperatures significantly higher than the corresponding equilibrium $T_\textrm{c}$.

To date, understanding the microscopic mechanism of photo-induced superconductivity remains an outstanding and active challenge.  
Two main classes of proposals have been given. In the first, the laser pulse serves as a quench of the system's parameters, switching the material into a distinct quasi-static state~\cite{PhysRevB.91.094308,PhysRevB.93.144506,lemonik2019transport,kennes2017transient} with, e.g.~a different lattice structure or interaction that leads to an increased equilibrium $T_\textrm{c}$. 
However, many of these proposals rely upon nonlinear phonon couplings, which predict a dependence of the conductivity on the drive amplitude inconsistent with further experiments~\cite{budden2021evidence,von2019parametrically}. 

In the second class, the laser pulse instead generates a coherent drive which creates an intrinsically non-equilibrium state. 
For example, in some of these materials, there is evidence that the laser light populates a  phonon mode, which subsequently drives the electronic degrees of freedom. 
During the ring-down of the phonon mode, the electrons are approximately governed by a time-periodic \textit{Floquet Hamiltonian}, 
$H(t) = H(t + 2\pi/\omega_\textrm{D})$, where $\omega_\textrm{D}$ is the phonon frequency. 
Within this framework, previous works have interpreted photo-induced superconductivity as a parametric resonance of emergent bosonic excitations, for example, uncondensed Cooper pairs, Josephson plasmons, the Higgs mode, or a fractionalized ``chargon'' excitation~\cite{PhysRevX.11.011055,PhysRevB.104.054512,PhysRevB.104.L241112,von2019parametrically,kleiner2020space,PhysRevB.103.224503,PhysRevLett.117.227001,PhysRevB.102.174505,komnik2016bcs,PhysRevB.96.045125,PhysRevB.96.144505,PhysRevB.94.214504,PhysRevB.96.014512}.

In the Cooper pair case~\cite{PhysRevB.104.054512}, electrons are assumed to pair at a relatively high temperature scale; however, in equilibrium, the coherence required for superconductivity may only occur at a much lower temperature. 
Thus, in the absence of a drive, the bosons would be incoherent. 
But when excited by the parametric resonance their population grows exponentially.
When interactions are ignored, so that each bosonic $k$-mode can be treated in isolation, the nature of this parametric resonance is simple and well-understood.
While this non-interacting parametric instability does not give rise to off-diagonal long-range order in space, which is a pre-requisite for a Meissner effect, it  nevertheless predicts the observed superconducting optical conductivity~\cite{PhysRevB.104.054512,PhysRevB.102.174505}

As the parametric resonance causes the boson populations to increase, we expect interactions to eventually become important, leading to coupling between the bosonic degrees of freedom. 
This leads us to two central questions: First, as the duration of the drive is increased, how does the  non-interacting parametric instability evolve into  an interacting non-equilibrium, many-body steady state? And second, would the interacting steady-state exhibit long-range coherence and a Meissner effect?

%Can such a coherent supercurrent exist in noisy driven systems, in the absence of a steady condensate?
%This question may sound far-fetched, but 
%One of the most convincing experimental signature is a $1/\omega$ imaginary part of the conductivity, resembling the London equation, down to the lowest frequency measured.

Interestingly, these questions connect photo-induced superconductivity with a broader class of ideas that fall under the umbrella of Floquet engineering---the use of pulsed-periodic control to modify the effective equations of motion of a many-body system.
Perhaps closest within this Floquet engineering landscape is the idea that periodically-driven systems can host intrinsically non-equilibrium phases of matter, such as the discrete time crystal (DTC)~\cite{bennett1990stability,gambetta2019classical,yao2020classical,PhysRevLett.117.090402,PhysRevLett.116.250401,PhysRevLett.118.030401,else2020discrete}.  
%Remarkably, recent theoretical work on DTCs addresses exactly the same type of questions raised above.
Of particular relevance is the theory of classical activated discrete time crystals~\cite{bennett1990stability,gambetta2019classical,yao2020classical}, which emerge in periodically driven arrays of non-linear oscillators coupled to a  bath. 
The resulting steady state exhibits sub-harmonic oscillations which spontaneously break time-translation symmetry up to an exponentially long time scale. Unlike previous proposals for DTCs in closed quantum systems~\cite{PhysRevLett.117.090402,PhysRevLett.116.250401,PhysRevLett.118.030401,PhysRevLett.115.256803,weidinger2017floquet,PhysRevX.7.011026,PhysRevX.10.011043}, an activated DTC exists in an open system in contact with a \textit{heat bath}, a situation  which is unavoidable in condensed matter experiments.  
%For a range of drive frequencies, dissipation strengths, and temperature, the oscillators settle into a steady state whose period is twice the period of the drive. 
%Surprising Even in 1D, non-equilibrium phase transition between disordered and period-doubled states with exponentially-long correlation time even in one dimension, in stark contrast to the expectations equilibrium statistical mechanics~\cite{yao2020classical}. 

In this work, we show that a parametrically driven many-body system with a global $U(1)$ symmetry can exhibit an intrinsically nonequilibrium phase corresponding to a period-doubled superconductor (PDSC).
The PDSC spontaneously breaks both the discrete time translation symmetry and the $U(1)$ symmetry, leading to a new example of a time crystal with a lifetime that we argue is infinite in three dimensions.  
Under a periodic drive of frequency $\omega_D$, the $U(1)$-breaking order parameter oscillates as %$ \psi(\mathbf{x}, t) \sim u\cos(\omega_Dt/2 + \theta_1)+i v\cos(\omega_Dt/2 + \theta_2)$.
$\psi(t, \mathbf{x}) \sim e^{i\phi}\cos(\omega_Dt/2 + \theta)$.
In contrast to scenarios where oscillations ``piggy-back'' off the $U(1)$ symmetry~\cite{else2020discrete},  when the $U(1)$ symmetry is explicitly broken, time-translation symmetry breaking will nevertheless persist out to an exponentially long time-scale, $\tau \sim e^{\Delta / T}$, where $T$ is the temperature of the bath. 

The interplay between the $U(1)$ phase symmetry and the discrete time-translation symmetry leads to a rich phase diagram (Figure 1),
%The real and imaginary component of $\psi$ can be understood as two parametrically driven oscillators; the nonlinear coupling between them leads to a much richer phase diagram than the activated discrete time crystal with a single component~\cite{yao2020classical}.
which we analyze by developing an effective field theory for the steady state. We show that when coupled to an electromagnetic field, the PDSC steady state exhibits a superconducting electromagnetic response with both perfect conductivity and the Meissner effect, even though the order parameter has zero time average due to its sub-harmonic oscillations.
We verify these  predictions via extensive numerical simulations of the Langevin dynamics of a bosonic array coupled to a coherent drive and a heat bath.

%Dissipation of particle and energy to the bath leads to rich nonequilibrium physics. We show that the dissipation modifies the low-energy spectrum, but the superconducting response remains unaffected by small dissipation. We derive a zero-temperature phase diagram in terms of the frequency of the drive and the strength the dissipation (relative to the strength of the drive). 

We next turn to a prediction of the PDSC transition temperature $T_\textrm{c}$. 
%We point out that although a general activated discrete time crystal, which by definition don't have any symmetry except time translation, have at most an exponentially long correlation time at finite temperature~\cite{yao2020classical}, 
In particular, we argue that the combination of the coherent Floquet drive and  the $U(1)$ symmetry can stabilize true long-range order in space and time at finite temperature. %This is contrast to the parametrically driven of a \emph{real} order parameter, which was found to give rise 
%to finite but thermally activated  length and time scales
%$\xi, \tau \sim e^{\DeltaT}$\cite{},  here w
%It is then interesting to investigate the transition temperature of this nonequilibrium superconductor. 
Guided by the effective theory, we further conjecture that the transition temperature is proportional to the strength of the periodic drive.
Thus in principle, $T_\textrm{c}$ can exceed the equilibrium transition temperature if the proper drive is chosen. %Experimental evidence for  transient superconductivity at the order of room temperature shows promise~\cite{cavalleri2018photo}
Finally, we propose a Josephson-tunneling experiment to detect the period-doubled oscillations of the superconducting order parameter. We note that our analysis also applies to periodically driven bosonic Mott insulators in cold atomic systems~\cite{zenesini2009coherent,eckardt2017colloquium}.

\section{Periodically-driven Bose Hubbard Model}

Let us begin by considering a lattice model and the corresponding continuum Hamiltonian of periodically-driven charged particles. We will start by connecting this model to the physics of an array of parametrically driven nonlinear oscillators~\cite{yao2020classical}.

\begin{figure}
\begin{center}
\includegraphics[width=0.4\textwidth]{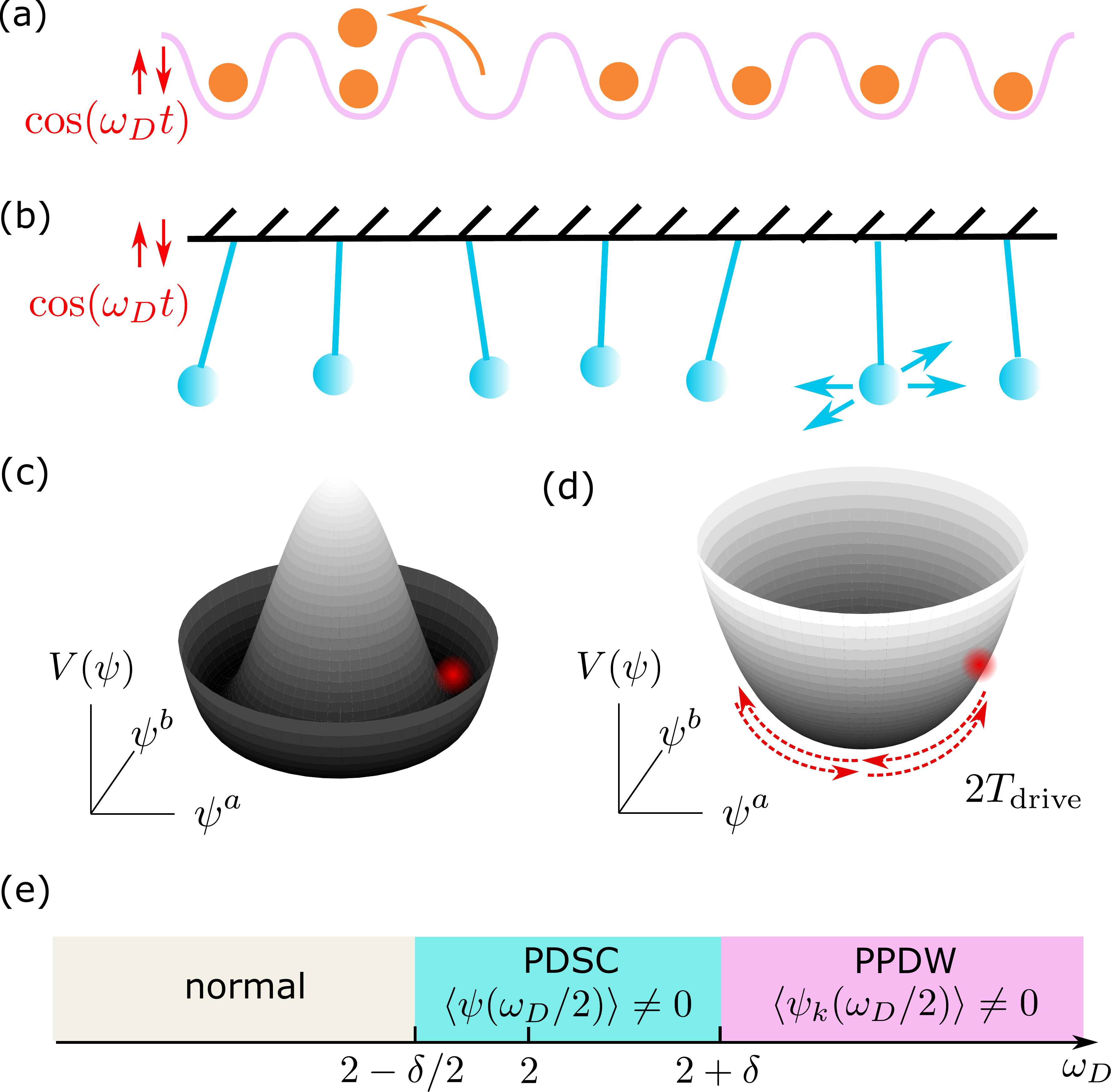}
\caption{(a) Illustration of the Mott phase of the Bose Hubbard model. Periodic modulation of either the hopping or the on-site repulsion creates particle-hole excitations. (b) Coupled pendula with two degrees of freedom (blue arrows) are described by the same continuum field theory as the boson model. (c) Mexican-hat potential (negative $\Delta$) in the equilibrium superconducting phase. The field picks a nonzero average value hence breaking the $U(1)$ symmetry (d) Potential energy for positive $\Delta$. Although the potential minimum is at $\psi^a=\psi^b=0$ at any instant, in the PDSC phase, the periodic drive induces a linear motion around the origin which breaks the $U(1)$ symmetry. (e) Phase diagram of the driven Bose Hubbard model at $\Delta = 1$. From left (low driving frequency) to right (high driving frequency): normal insulating phase that preserves the discrete time-translation symmetry; period-doubled superconducting phase (PDSC) that breaks the U(1) symmetry and the discrete time-translation symmetry; period-doubled pair density wave (PPDW) phase that breaks the U(1) symmetry, the discrete time-translation symmetry and spatial translation symmetry. In the main text, we introduce a dimensionless measure of the detuning $u = 4(1-\omega_D/2)/\delta$. The PDSC phase appears at $-2<u<1$.}
\label{Figcartoon}
\end{center}
\end{figure}

One of the simplest models that exhibits photo-induced superconductivity is the periodically-driven Bose Hubbard model at \textit{integer} filling~\cite{PhysRevB.104.054512,PhysRevB.104.L241112}:
\begin{align}
H = &-(t+\delta t\cos(\omega_D t))\sum_{\<ij\>}(b_i^\dagger b_j + h.c.) \nonumber\\
&+ (U + \delta U\cos(\omega_Dt))\sum_{i} (n_i-\bar{n})^2,
\end{align}
where $\omega_D$ is the frequency of the drive, $b_i$ is the boson annihilation operator at site $i$, $n_i$ is the boson number at site i, and $\bar{n}=1$ is the filling. Consider the insulating phase, where the repulsive Hubbard interaction is larger than the bandwidth. Roughly speaking, the ground state is captured by the simple cartoon picture of one boson per site. The insulating gap depends on both the on-site repulsion and the hopping, and  periodic modulations of these parameters create particle-hole excitations above the ground state [Fig.~\ref{Figcartoon}(a)]. 

To  describe the  system, we use a continuum theory of a complex boson field, $\psi$. It is well known that a relativistic bosonic field theory~\cite{PhysRevB.40.546} describes the low energy physics of the Hubbard model at an integer filling~\footnote{Away from an integer filling, there is a chemical potential term $\psi\partial_t\psi$ which makes the low-energy dispersion non-relativistic.}.
The periodic drive gives time dependence to every coefficient of the field theory. The simplest symmetry allowed term is a coupling to $|\psi|^2$, hence the continuum Hamiltonian~\cite{PhysRevB.104.054512},
\begin{align}
H_\text{con} = \int d^dx  [&\frac{1}{2}|\pi|^2 + \frac{g}{2}|\mathbf{\nabla}\psi|^2 +V(\psi) +\frac{\delta}{2}\cos(\omega_D t)|\psi|^2],\nonumber\\
  &\textrm{with} \hspace{2mm} V(\psi) =  \frac{\Delta}{2}|\psi|^2 -\frac{\epsilon}{3}|\psi|^4.
\label{EqHamiltonian}
\end{align}
The potential energy, $V(\psi)$, takes the standard Ginzburg-Landau form, $\delta$ is the effective strength of the periodic drive,  $\pi\equiv\partial_t\psi$ is the  canonical momentum conjugate to $\psi$, and for the Bose Hubbard model, $\epsilon < 0$, so that the potential is bounded  from below. 
The global $U(1)$ symmetry acts as $\psi\rightarrow e^{i\phi_0}\psi$ and we define $\psi^a=\text{Re}\psi, \psi^b=\text{Im}\psi, \pi^a=\text{Re}\pi, \pi^b=\text{Im}\pi$, with
\begin{align}
    &[\psi^\sigma(x),\pi^{\sigma'}(x')]= i\delta_{\sigma\sigma'}\delta(x-x'),&\sigma,\sigma'\in\{a,b\}
    \label{Eqcanonicalcommutation}
\end{align}
At early times, when the interactions between the excitations can be neglected, one can utilize Eq.~\ref{EqHamiltonian} to study the quantum state of the bosons after turning on the periodic drive~\cite{PhysRevB.104.054512}.
In this case, the system is unstable to a parametric resonance within a momentum shell  which is centered on an energy of half the drive frequency, $E_k \sim \hbar \omega_D / 2$, and with a width in proportion to the driving amplitude $\delta$.
The resulting transient state exhibits a superconducting-like AC response but no Meissner effect~\cite{PhysRevB.104.054512}. 
In this work, we demonstrate that this transient state eventually settles into a steady state where a true condensate forms due to interactions and dissipation. The oscillations of the boson field in the steady state resemble that of an activated discrete time crystal found in driven arrays of classical pendula [Fig.~\ref{Figcartoon}(b)]~\cite{yao2020classical}. We can view $\psi^{a,b}$ and $\pi^{a,b}$ as the positions and momenta of the pendula (with generalized potential energy $V(\psi)$) in the $a$ and $b$ directions respectively, and $H_\text{con}$ as the Hamiltonian of the coupled pendula system. As we will see, this point of view is particularly useful when describing an ordered state.

Let us now briefly discuss the equilibrium and non-equilibrium physics using the continuum Hamiltonian. In equilibrium ($\delta=0$), the ground state is determined by the sign of $\Delta$. For $\Delta>0$, one has the insulating phase, where $\<\psi\>=0$ in the ground state and the  insulating gap is $\sqrt{\Delta}$ (at the mean field level).
For $\Delta<0$, one has  the superfluid phase, where the potential has the shape of a Mexican hat [Fig.~\ref{Figcartoon}(c)]. In the ground state, the boson field picks a minimum, $\<\psi\>\neq 0$, spontaneously breaking the $U(1)$ symmetry. 
%The equilibrium physics of the Bose Hubbard model is well understood, but the periodic drive brings rich physics that has not been studied to date.
Interestingly, we will show that 
 with a periodic drive, a new superconducting steady state exists for $\Delta>0$. The intuition is the following: even though the potential minimum is at $\psi=0$ at any instant in time, the periodic drive can induce a stable motion along some axis in the a-b plane, as depicted in Fig.~\ref{Figcartoon}(d). This motion picks a specific axis and therefore it spontaneously breaks the $U(1)$ symmetry. Crucially, $\<\psi(t)\>$ exhibits a period which is twice the period of the underlying drive in the steady state; this is in direct analogy to the dynamics of parametrically driven oscillators in the so-called activated discrete time crystalline phase~\cite{yao2020classical}, hence the name period-doubled superconductivity (PDSC).

To analyze the late-time behavior of the driven boson system, we take the following strategy. We first assume that the parametrically excited particles and holes settle into a condensate, whose dynamics are characterized by the equations of motion of $\<\psi(\vec{r},t)\>$.
By investigating this equation of motion, we will derive an effective theory of the non-equilibrium steady state, and then discuss the stability of this state.

Starting from the insulating phase, $\Delta>0$  (hereon, we choose units in which $\Delta=1$), and  following Eq.~(\ref{EqHamiltonian}-\ref{Eqcanonicalcommutation}), one finds that the
 equation of motion for the field $\psi$ is given by:
\begin{align}
    \partial^2_t\psi = -\psi + \frac{4\epsilon}{3}|\psi|^2\psi - g\mathbf{\nabla}^2\psi -\delta\cos(\omega_D t)\psi
    \label{EqEOM}
\end{align}
Such dynamics are well-known to exhibit a parametric resonance when $\omega_D\simeq 2$. 
Focusing on the uniform ansatz $\psi(t, x) = e^{i \phi} q(t)$, where $q$ is real, the equations of motion reduce to a single parametrically driven oscillator, $\ddot{q} = -(1 + \delta \cos(\omega_D t))q + \frac{4 \epsilon}{3} q^3$.
With our choice of units $q$ has a natural frequency of unity, so when $\omega_D \sim 2$ the term $\delta\cos(\omega_D t) q$ contains a Fourier component at frequency $\omega_D-1\simeq 1$, which resonantly contributes to the motion of $q$ (Eq.~\ref{EqEOM}). 
This is the origin of the parametric resonance. 
Without the nonlinear term, $q \sim \exp(\lambda t)\cos(\omega_Dt/2+\theta_0)$, grows exponentially when $|\omega_D/2-1|<\delta/4$. The nonlinear term eventually stops the exponential growth at late times and (with infinitesimal dissipation) pins the oscillation frequency to exactly $\omega_D/2$.

For the periodically driven Bose-Hubbard model,  $\psi^a\equiv \text{Re}\psi$ and $\psi^b\equiv \text{Im}\psi$ yield two degrees of freedom related by the $U(1)$ symmetry. Solutions of the form 
 $\psi(t,x) = e^{i \phi} q(t) \sim \psi_0\cos(\omega_D t/2 +\theta_0)$ pick a particular axis in this space, and hence break the $U(1)$ symmetry. We find solutions of this form are stable in a frequency range around $\omega_D/2 \sim 1$ (Fig.~\ref{Figcartoon}(e)). For larger frequencies, other types of period-doubled superconducting states which break spatial translation symmetry instead become stable.
We denote these states as   period-doubled pair density waves (PPDW). 

%Recently the original studies of period doubling in classical dynamical systems are rejuvenated by the characterization of non-equilibrium many-body phases of matter. Ref... showed that the driven FK model in contact with a low-temperature heat bath realizes an activated classical discrete time crystal, which is robust up to an exponentially long time scale in the limit $\delta,\omega_D/2-1\ll\omega_D$. This inspires us to consider potential time crystal phases in solid state materials. 
%We will show that the solution is stable for a range of tuning parameters; it gives an intrinsically nonequilibrium superconductor, exhibiting perfect conductivity and Meissner effect. The stability of the state is modified by the extra degrees of freedom and the presence of the $U(1)$ symmetry in a nontrivial way. We expect a true time crystal that is stable for \textit{infinitely} long time even though the original driven FK model only realized an exponentially long correlation time. 

\section{Analysis of the period-doubled steady state}
\label{SecSteadyState}

We now turn to analyzing the steady state by deriving an effective time-independent Hamiltonian, which governs the dynamics in the rotating frame of the parametric resonance. 
In short, we first use a time-dependent canonical transformation to enter a rotating frame in which the new canonical variables,  $(\tilde{\psi}^\sigma,\tilde{\pi}^{\sigma'})$, will vary slowly at resonance. 
We then obtain an effective time-independent Hamiltonian for their slow dynamics via a lowest-order Magnus expansion.
Finally, we introduce a new set of canonical variables $(J,\theta,n,\phi)$ which will provide a more convenient representation of the $U(1)$ symmetry. 
%
%Readers happy to skip the technical details of these transformations can proceed to Eq.~\ref{EqHbarJnthetaphi}.
We will then use this set of  canonical variables to solve for the steady state, the excitation spectrum, the conditions under which the steady state is stable, as well as to ultimately connect the effective theory of the nonequilibrium steady state to that of a conventional superconductor.

\begin{figure}[htb]
\begin{center}
\includegraphics[width=0.4\textwidth]{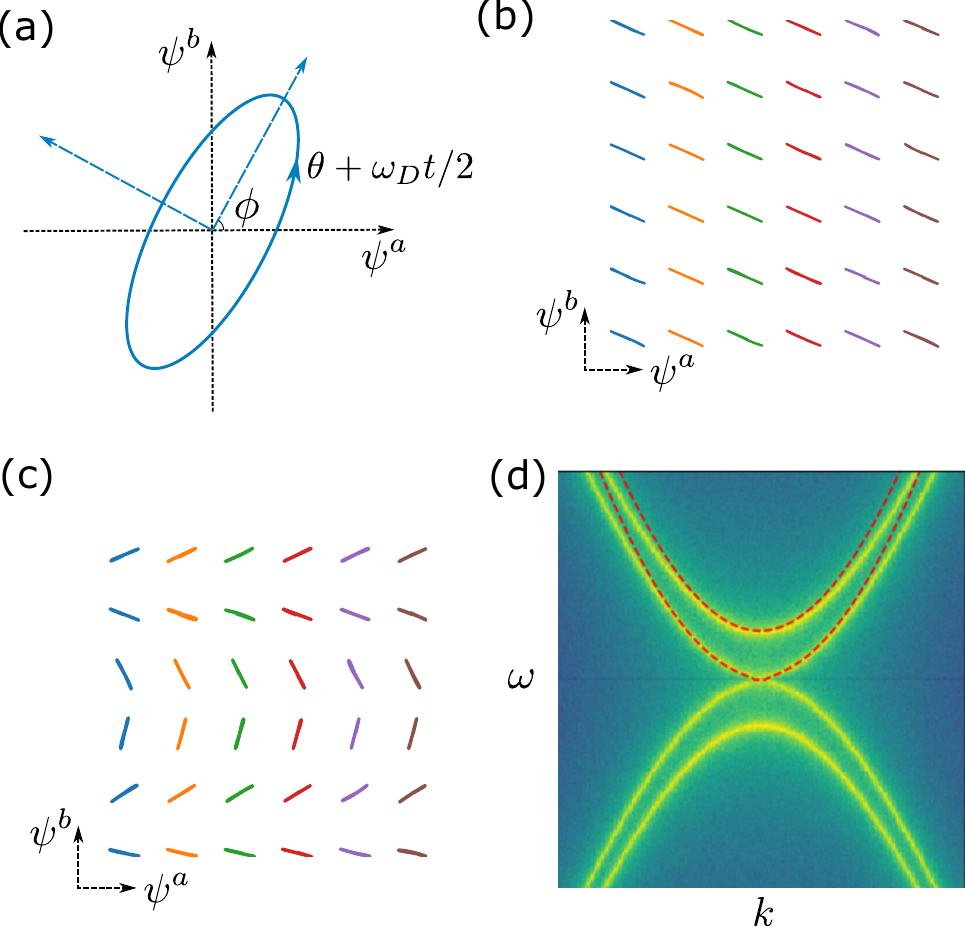}
\caption{(a) Illustration of the time evolution of the boson field $\psi^\sigma$ given a fixed set of $(J,\theta,n,\phi)$. The new canonical variables $J$, $\theta$, $n$, and $\phi$ represent the slowly changing size, phase shift, shape, and polarization angle of an elliptic orbit in the $\psi^a$-$\psi^b$ plane. The elliptic orbit degenerate into a linear orbit when $n=0$. (b) Steady state orbits at each spatial point in PDSC. $u=-1, \delta=0.1, g=1, \eta = 0.01, \text{T}_\text{eff} = 0.0001$, showing a 6 by 6 patch of a 50 by 50 lattice. See Appendix~\ref{AppendixNumerics} for details of the simulation.
(c) The same as (b) except for the PPDW state at $u=-10$. (d) Stroboscopic spectral function $S(\omega,k)=\<|\text{DFT}(-1)^n\psi_j(nt_D)|^2\>$ for the PDSC state, measured at $u=-1, \delta=0.1,g=1, \eta=0.005, \text{T}_\text{eff}=0.00001$ on a 1D lattice with 600 sites. The period-doubled motion is mapped to $\omega=k=0$. The noise spectrum fits well with the theoretical predictions (dashed red lines) of the generalized Higgs mode and the Goldstone mode at small frequencies.}
\label{FigOrbitSpectrum}
\end{center}
\end{figure}

First, we make the following time-dependent canonical transformation $(\psi^\sigma,\pi^{\sigma'},H_\text{con}(t))\rightarrow (\tilde{\psi}^\sigma,\tilde{\pi}^{\sigma'},\tilde{H}_\text{con}(t))$:
\begin{align}
\left\{ \begin{array}{l}
\psi^\sigma = \tilde{\psi}^\sigma \cos(\omega_Dt/2) + \tilde{\pi}^\sigma \sin(\omega_Dt/2)\\
\pi^{\sigma'} = \tilde{\pi}^{\sigma'} \cos(\omega_Dt/2) - \tilde{\psi}^{\sigma'} \sin(\omega_Dt/2)\\
H_\text{con}(t) = \tilde{H}_\text{con}(t) + \int d^dx \ \frac{\omega_D}{4}(|\pi|^2 + |\psi|^2).
\end{array} \right.
\label{Eqcanonicaltransformation}
\end{align}
We are interested in the scenario where $\psi$ and $\pi$ oscillate at frequencies near $\omega_D/2$. Thus, the new variables $\tilde{\psi}^\sigma$ and $\tilde{\pi}^{\sigma'}$ will vary slowly and will characterize the  shape and phase of the elliptic orbit of the complex boson field [Fig.~\ref{FigOrbitSpectrum}(a)]. $\tilde{H}_\text{con}$ still contains oscillatory terms at frequency $\omega_D$, but these fast oscillations will have only a small average effect on the slowly varying $\tilde{\psi}^\sigma$ and $\tilde{\pi}^{\sigma'}$. We can thus utilize the Floquet-Magnus expansion to obtain a \textit{static} effective Hamiltonian. To  leading order in $\delta/\omega_D$, the time-independent Hamiltonian is just the time-average of $\tilde{H}_\text{con}$:
\begin{align}
H_\text{eff} = \frac{1}{t_D}\int_0^{t_D}\tilde{H}_\text{con}(\tilde{\psi},\tilde{\pi},t)dt
\end{align}
\begin{align}
=\int \ [&\frac{\delta u}{8}(|\tilde{\psi}|^2 + |\tilde{\pi}|^2) + \frac{\delta}{8}(|\tilde{\psi}|^2- |\tilde{\pi}|^2) +\frac{g}{4}(|\mathbf{\nabla}\tilde{\psi}|^2+|\mathbf{\nabla}\tilde{\pi}|^2)\nonumber\\ &- \frac{\epsilon}{8}(|\tilde{\psi}|^2 +|\tilde{\pi}|^2)^2
+ \frac{\epsilon}{6}(\tilde{\psi}^a\tilde{\pi}^b - \tilde{\psi}^b\tilde{\pi}^a)^2\ ]d^dx
\label{EqHbarpsipi}
\end{align}
\begin{align}
&\partial_t\tilde{\psi}^\sigma(x) = \frac{\delta H_\text{eff}}{\delta \tilde{\pi}^{\sigma}(x)}, &\partial_t\tilde{\pi}^{\sigma}(x) = -\frac{\delta H_\text{eff}}{\delta \tilde{\psi}^{\sigma}(x)}
\label{EqaveEOM}
\end{align}
where $u=4(1-\omega_D/2)/\delta$ is a measure of the detuning. %
%\mz{I did not check the coefficients myself :)}
%
One can directly find `stationary points', $\partial_t\tilde{\psi}^a=\partial_t\tilde{\psi}^b=\partial_t\tilde{\pi}^a=\partial_t\tilde{\pi}^b=0$, from Eq.~\ref{EqaveEOM}. By definition, stationary points in the rotating frame correspond to period-doubled steady states in the original frame.

To better represent the $U(1)$ symmetry and the discrete time translation symmetry, we make a further canonical transformation. We change the canonical variables  from $(\tilde{\pi}^a,\tilde{\psi}^a,\tilde{\pi}^b,\tilde{\psi}^b)$ to $(J,\theta,n,\phi)$ such that $(\tilde{\psi}^a,\tilde{\psi}^b)^\text{T} = R(\phi)(\sqrt{2J_1}\cos(\theta),\sqrt{2J_2}\sin(\theta))^\text{T}$ and $(\tilde{\pi}^a,\tilde{\pi}^b)^\text{T} = R(\phi)(-\sqrt{2J_1}\sin(\theta),\sqrt{2J_2}\cos(\theta))^\text{T}$, where
\begin{align}
R(\phi) =
\left(\begin{array}{cc}
\cos(\phi) & -\sin(\phi)\\
\sin(\phi) & \cos(\phi)
\end{array}\right), J_{1,2} = (J \pm \sqrt{J^2-n^2})/2.
\end{align}
The new variables $J$ and $\theta$ are conjugate to each other, $[J(x),\theta(x')] =-i\delta(x-x')$,   $J=(|\tilde{\psi}|^2+|\tilde{\pi}|^2)/2=(|\psi|^2+|\pi|^2)/2$ is the quadratic part of the energy density of the oscillators and $\theta$ is the phase shift of the period-doubled motion. 
Time translation by one driving period acts as $\theta\rightarrow\theta+\pi$, while $n$ and $\phi$ are the conjugate momentum and coordinate that represent the degree of freedom of rotation in the a-b plane, $[n(x),\phi(x')] =-i\delta(x-x')$. $n=\tilde{\psi}^a\tilde{\pi}^b-\tilde{\psi}^b\tilde{\pi}^a = \psi^a\pi^b-\psi^b\pi^a$ is the angular momentum density of the oscillators, which corresponds to the charge density of the boson relative to an integer filling (we will show this later explicitly by minimally coupling the boson model to an electromagnetic field); $\phi$ is the polar angle of the orbit in the a-b plane and the $U(1)$ symmetry acts as $\phi\rightarrow \phi + \phi_0$.

The average Hamiltonian, $H_\text{eff}$, becomes
\begin{align}
    H_\text{eff} = \int d^dx[\frac{\delta u}{4}J -\frac{\epsilon}{2}J^2 + \frac{\epsilon}{6}n^2+\frac{\delta}{4}\sqrt{J^2-n^2}\cos(2\theta)\dots],
    \label{EqHbarJnthetaphi}
\end{align}
where we omit, for now, the terms involving spatial derivatives. Note that $H_\text{eff}$ is invariant under both the $U(1)$ symmetry, $\phi\rightarrow \phi + \phi_0$ and the discrete time translation symmetry, $\theta\rightarrow\theta+\pi$. We  will focus on uniform steady states with  uni-axial motion, namely $n=0$, which describe uniform states at \textit{integer} filling of the periodically driven Bose-Hubbard model. For $\epsilon<0$, there is only one such steady state up to the $U(1)$ rotation and the discrete time translation, sitting at the minimum of $H_\text{eff}$: $J_0=\delta (u-1)/(4\epsilon)$, $n_0=0$, $\phi=\text{const.}$, and $\theta_0=\pi/2$. In the steady state, the original boson field oscillates as $\psi\simeq \sqrt{2J_0}\sin(\omega_Dt/2)e^{i\phi}$ [Fig.~\ref{FigOrbitSpectrum}(b)].

\subsection{Excitation spectrum and stability condition}
We pause here to compare the steady state we just found with the parametric resonance of the free boson model (which describes the transient state shortly after the pump~\cite{PhysRevB.104.054512}). 
For free bosons, each wavevector hosts independent modes; 
parametric resonances set up at all wavevectors where $|2E_k -\omega_D| < \delta/2$, and there is no long-range order in space.
Interactions, however, lead to a condensate at a single momentum (or a few discrete momenta), determined by the drive frequency.

In this section, we work out the frequency range where the uniform steady state is stable and calculate the excitation spectrum above the steady state. To do so, we expand $H_\text{eff}$ to the lowest order in $\tilde{J}=J-J_0$ and $n$:

\begin{align}
    H_\text{eff} \simeq \int& d^dx [-\frac{\epsilon}{2}\tilde{J}^2 + \frac{a}{4}(\cos(2\theta)+1) + \frac{b}{2}n^2\nonumber\\
    &+\frac{gJ_0}{2}(|\mathbf{\nabla}\theta|^2 +|\mathbf{\nabla}\phi|^2) + \frac{g}{8J_0}(|\mathbf{\nabla}n|^2 +|\mathbf{\nabla}\tilde{J}|^2)]
    \label{EqHbarexpansion}
\end{align}
where $a=\delta^2(u-1)/(4\epsilon), b=\epsilon(u+2)/(3u-3)$. Recall that $\epsilon <0$. The steady state can be stable only when $a,b>0$, otherwise small deviations from the steady state grow exponentially over time. In terms of the detuning, the stability condition is \mbox{$-2<u<1$} (Fig.~\ref{Figcartoon}(e)). For $u<-2$, namely when $\omega_D$ is considerably larger than twice the boson gap, we numerically find that the uniform state eventually evolves to various period-doubled translation-symmetry-breaking superconducting states (PPDW), one of which has $\psi\simeq\sqrt{2J_0}\sin(\omega_Dt/2)e^{i(\phi + kx)}$ (Fig.~\ref{FigOrbitSpectrum}(c)). In these states, the bosons condense at the momentum $k$ where $\omega_D\simeq 2E_k$.

In the uniform steady state, we see from Eq.~\ref{EqHbarexpansion} that the conjugate variables $\tilde{J}$ and $\theta$ give a gapped mode, with dispersion $\omega_\text{H} \simeq \sqrt{(-\epsilon+gk^2/(4J_0))(a+gJ_0k^2)}$. This mode is analogous to the Higgs mode of an equilibrium superconductor since $\tilde{J}$ represents fluctuations of the oscillation amplitude of the boson field. On the other hand, $n$ and $\phi$ give a gapless mode, with linear dispersion near zero momentum, $\omega_{G}\simeq \sqrt{gJ_0k^2(b+gk^2/(4J_0))}$, which we identify as the Goldstone mode. We remind the readers that these expressions of the dispersion relation are only accurate when $\omega_G,\omega_H\ll \omega_D$, due to the approximations we made in the Floquet-Magnus expansion. 

We numerically confirm the existence of these modes (Fig.~\ref{FigOrbitSpectrum}(d)).
We add a small dissipation term $-\eta\partial_t\psi$ to the right hand side of the equation of motion in the original frame, Eq.~\ref{EqEOM}, to eliminate transient behaviors. We then add a small amount of Gaussian noise, $\xi(x,t)$,  $\<\xi^\sigma(x,t)\xi^{\sigma'}(x',t')\>=2\eta T_\text{eff}\delta^{\sigma,\sigma'}\delta(x-x')\delta(t-t')$ and compute the stroboscopic spectral function of the fluctuations above the steady state, $S(\omega,k)=\<|\sum e^{-ikj + i\omega nt_D}(-1)^n\psi_j(nt_D)|^2\>$.
$S(\omega, k)$ exhibits peaks in exact agreement with the predicted $\omega_{H/G}(k)$. 
See Sec.~\ref{subsecDissipation} and Appendix~\ref{AppendixNumerics} for details of the simulation.

\subsection{Electromagnetic response of the steady state}

We now analyze the electromagnetic response of the steady state. We minimally couple the boson model (Eq.~\ref{EqHamiltonian}) to an electromagnetic field: $\mathbf{\nabla}\psi\rightarrow(\mathbf{\nabla}+i\vec{A})\psi$, $\partial_t\psi\rightarrow(\partial_t+iA_0)\psi$. Thus, the canonical momentum changes to $\pi^a=\partial_t\psi^a-A_0\psi^b,\pi^b=\partial_t\psi^b+A_0\psi^a$, and the continuum Hamiltonian becomes

\begin{align}
H_\text{con}(A) = &\int d^dx[\frac{1}{2}|\pi|^2  +V(\psi) +\frac{\delta}{2}\cos(\omega_D t)|\psi|^2\nonumber\\
&+\frac{g}{2}|(\mathbf{\nabla}+i\vec{A})\psi|^2 -A_0(\psi^a\pi^b-\psi^b\pi^a)].
\label{EqHamiltonianApsi}
\end{align}
The current density is
\begin{align}
    \vec{j}=-\frac{\delta H_\text{con}(A)}{\delta\vec{A}} = (-i\psi^*\mathbf{\nabla}\psi + h.c) -g|\psi|^2\cdot \vec{A}.
\end{align}

To derive an effective theory for the low-frequency electromagnetic response, we use the canonical variables $(J,\theta,n,\phi)$ as before and apply the Floquet-Magnus expansion to $\tilde{H}_\text{con}(A)$. Assuming that $A$ is constant within a Floquet cycle, we find the effective Hamiltonian is modified by the substitutions
$\mathbf{\nabla}\phi\rightarrow\mathbf{\nabla}\phi + A$ and $H_\text{eff}\rightarrow H_\text{eff} -A_0n$ in Eq~\ref{EqHbarexpansion}. Furthermore, for low-frequency responses, we can ignore fluctuations of the gapped mode and set $J\simeq J_0,\theta\simeq\theta_0$; we can also ignore  the $|\mathbf{\nabla}n|^2$ in Eq.~\ref{EqHbarexpansion} since it involves more derivatives than the $n^2$ term. Thus, the low-frequency electromagnetic response is given by the following effective Hamiltonian

\begin{align}
    H_\text{eff}(n,\phi,A)\simeq \int d^dx\ [ \frac{b}{2}n^2-A_0n + \frac{gJ_0}{2}(\mathbf{\nabla}\phi+\vec{A})^2].
\end{align}
Remarkably, this effective Hamiltonian is the same as that of an equilibrium superconductor, with a superfluid density proportional to $J_0$. The current density is
\begin{align}
    \vec{j} = -gJ_0(\mathbf{\nabla}\phi + \vec{A}),
    \label{EqLondon1}
\end{align}
which is the same as the London equation. Thus, one finds that a period-doubled non-equilibrium superconductor emerges from the  driving of an insulator!

\begin{figure}[htb]
\begin{center}
\includegraphics[width=0.5\textwidth]{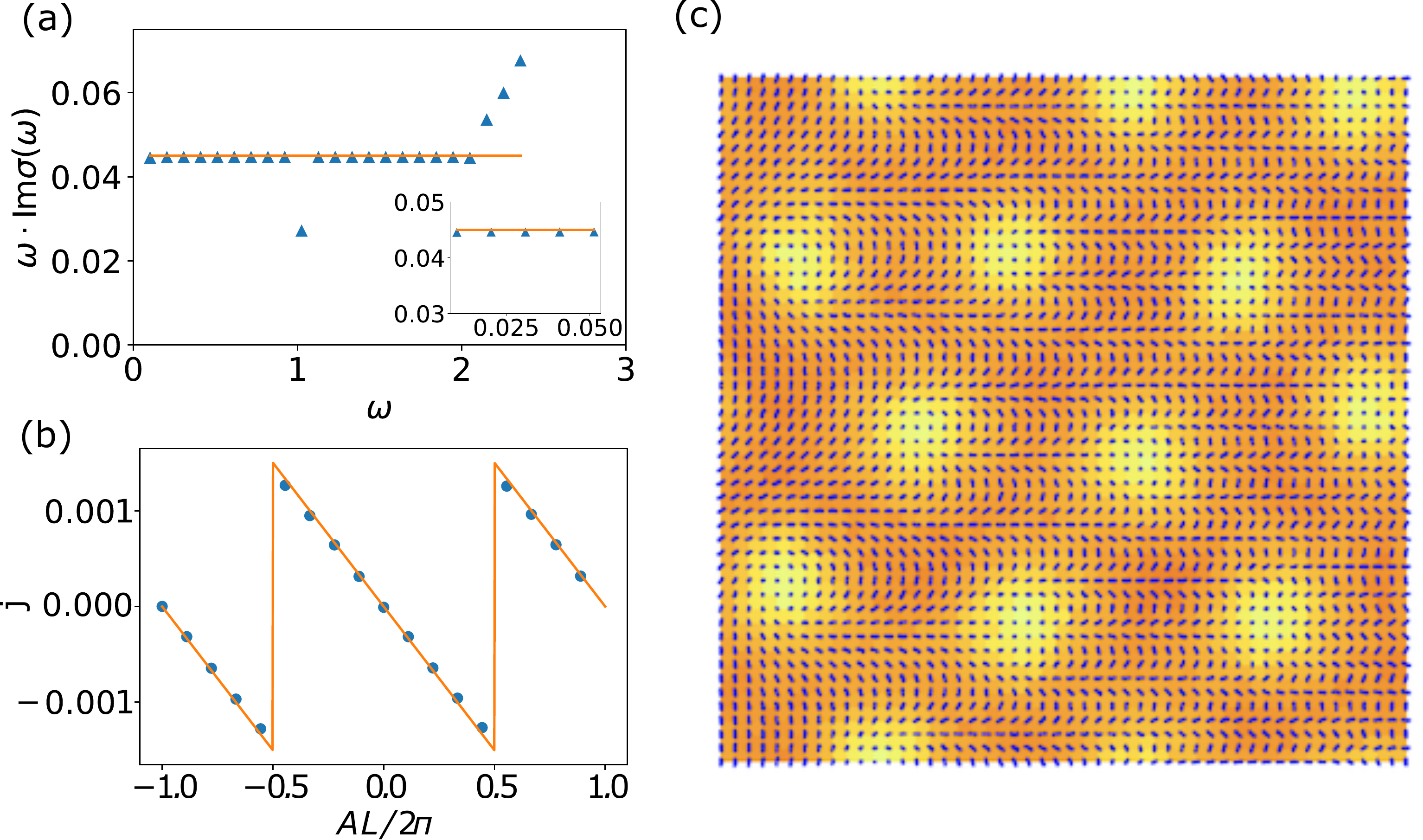}
\caption{(a) $\omega\text{Im}\sigma(\omega)$ for frequencies up to $1.2\omega_D$. $u=-1,\delta=0.1, g=1, \eta=0.03,\text{T}_\text{eff}=0.0001, L=200$. Simulation results (blue triangles) match the theoretical prediction for superfluid density (yellow line) at low frequencies (inset). (b) Steady-state current (blue dots) under a constant vector potential $A$. $u=-1,\delta=0.1, g=1, \eta=0.02, \text{T}_\text{eff}=0.003, L=100$. The yellow line shows the theoretical prediction. (c) Vortex lattice on torus with 12 flux. $u=-1,\delta=0.1, g=1, \eta=0.03,\text{T}_\text{eff}=0.003,  L=50$. The blue lines show the steady-state trajectories of $\psi^a-\psi^b$ at each spatial point. The color map show the corresponding oscillation amplitude. The center of each vortex where the amplitude goes to zero is shown as bright yellow in the colormap.}
\label{FigEMresponse}
\end{center}
\end{figure}

We check these theoretical predictions for the electromagnetic response by simulating the dynamics under the time-dependent Hamiltonian (Eq.~\ref{EqHamiltonianApsi}), with a small dissipation $\eta$, and an infinitesimal $T_\text{eff}$ as described in the last section, to remove transient behavior.
We first simulate (see Sec.~\ref{subsecDissipation} and Appendix~\ref{AppendixNumerics} for details) the response of PDSC to an AC electric field (Fig.~\ref{FigEMresponse}(a)).
We initialize the system in the uniform steady state, and then slowly turn on a uniform vector potential $\vec{A} =A_0f(t)\cos(\omega t)\hat{x}$, where $f(t)=0$ at $t=0$ and slowly grows to 1 after time $t_0$, after which we measure the current at frequency $\omega$ and compute the AC conductivity. We see that the theoretical prediction (yellow line) matches the numerical result at almost all frequencies smaller than $\omega_D$, down to the lowest frequencies tested. The prediction breaks down for a single point below $\omega_D$ at $\omega=\omega_D/2$ ($\omega= 1.025$ in Fig.~\ref{FigEMresponse}(a))  
due to interference with the oscillation of the order parameter and for $\omega >\omega_D$ due to resonant particle-hole creation.

We next simulate the response of the steady state on a torus to a static vector potential $\vec{A}=A\hat{x}$, which corresponds to a twist of the boundary condition by phase $AL/2\pi$.
Starting from the initial state with $\psi=0,\pi=0$ on a 2D lattice, we find that the final state minimizes $|\vec{j}|^2\propto|\mathbf{\nabla}\phi+\vec{A}|^2$ by choosing the winding number of the $U(1)$ phase (Fig.~\ref{FigEMresponse}(b)), which is well-known in equilibrium superconductors. When $|AL/2\pi|<0.5$, $\mathbf{\nabla}\phi=0$ and $\vec{j}=-gJ_0\vec{A}$. $\mathbf{\nabla}\phi$ jumps by $\pm2\pi/L$ at $AL/2\pi=\pm0.5$. Finally, we test the response to a uniform magnetic field. We simulate the equations of motion with 12 flux quanta on a torus (Appendix~\ref{AppendixNumerics}). The steady state shows a vortex lattice with 12 vortices as expected (Fig.~\ref{FigEMresponse}(c)).

\section{Dissipation and Noise}
\label{subsecDissipation}
In the previous sections, we have found an effective Hamiltonian by using canonical transformations and the Floquet-Magnus expansion. 
We have discussed the physical properties of the steady state at the \textit{minimum} of the effective Hamiltonian, but we have yet to discuss the role of the heat bath. 
Under what conditions of the system-bath coupling does the driven system approach the minimum of the effective Hamiltonian? 
This question may sound trivial at the first glance -- for an equilibrium system to approach its energy minimum, a low-temperature bath is all that is needed.
However, in the non-equilibrium setting, a low-temperature bath does not guarantee low entropy of the system.

In this section, we take the first step toward answering these questions. We first discuss the problem of a driven quantum system in contact with a heat bath in a general setting, and argue that under \textit{weak} system-bath coupling, the steady state of the system is \textit{approximately} a thermal state of the effective Hamiltonian. Nonetheless, the effective temperature is in general different from the temperature of the bath. 
We will discuss a special case when the two temperatures are equal. 
In the end, we introduce a nonequilibrium Langevin equation to describe the dynamics of the system in the presence of a bath.

Consider the following Hamiltonian of the system and the bath,
\begin{align}
    H = H_\text{sys}(t) + H_\text{bath} + H_I,
\end{align}
where $H_\text{sys}(t)$ is periodic and $H_\text{bath}$ and $H_I$ are independent of time. Following the preceding analysis, we can rewrite operators of the system in the rotating frame (as in Eq.~\ref{Eqcanonicaltransformation}) and replace the time-dependent Hamiltonian by the static Hamiltonian $H_\text{eff}$. Note that this causes the interaction term between the system and the bath to develop time dependence. Thus the total Hamiltonian becomes
\begin{align}
    H = H_\text{eff} + H_\text{bath} + \tilde{H}_I(t),\\
    \tilde{H}_I(t) = \tilde{H}_I(t+ 2T_D).
\end{align}
Due to this time-dependence, the ``total energy'' measured in the rotating frame is not conserved; therefore, the steady state of the system is in general not a thermal state of the effective Hamiltonian in the rotating frame. However, when the system-bath coupling is weak, the steady state of the system is locally equivalent to a thermal state of $H_\text{eff}$. This is because the system has enough time to thermalize to an effective temperature $T_\text{eff}$ between two consecutive perturbations from the bath. $T_\text{eff}$ is determined by details of the periodic driving and the coupling between the system and the bath.
%``energy density" of the system in the presence of energy exchange with the bath and extra energy change by the time-dependent $\tilde{H}_I$. 
In general $T_\text{eff}\neq T_\text{bath}$~\cite{kohn2001periodic,ikeda2020general}. In the special case when $\tilde{H}_I$ is dominated by its constant component, $T_\text{eff}\simeq T_\text{bath}$ in analogy to equilibrium thermodynamics. We discuss a natural system-bath interaction with this property in Appendix~\ref{AppendixPhononBath}.

In general, we can trace out the bath and derive a \textit{nonequilibrium} effective theory of the system alone. To do that, we borrow the wisdom of the exciton-polariton community ~\cite{PhysRevX.5.011017,PhysRevLett.97.236808,PhysRevA.82.043612,PhysRevB.85.184302,PhysRevA.87.023831,PhysRevLett.110.195301,PhysRevB.89.134310}, where non-equilibrium superfluids are discussed in detail. Starting from the Keldysh formalism, Langevin-type equations are derived which do not satisfy detailed balance. The effective noise level can be either stronger or weaker than the thermal noise depending on details of the coupling to the heat bath. 
In the longwavelength limit 
$\<\tilde{\xi}^\sigma(x,t)\tilde{\xi}^{\sigma'}(x',t')\>=2\eta T_\text{eff}\delta^{\sigma,\sigma'}\delta(x-x')\delta(t-t')$.
The simplest way to model the dissipation is to give the bosonic quasiparticles a decay rate, $\eta$. Together with the noise, we have the following equation of motion in the rotating frame:

\begin{align}
&\left\{ \begin{array}{l}
\partial_t\tilde{\psi}^\sigma(x) = \frac{\delta \tilde{H}_\text{con}}{\delta \tilde{\pi}^{\sigma}(x)} - \frac{\eta}{2}\tilde{\psi}^\sigma(x)\\ \partial_t\tilde{\pi}^{\sigma}(x) = -\frac{\delta \tilde{H}_\text{con}}{\delta \tilde{\psi}^{\sigma}(x)}-\frac{\eta}{2}\tilde{\pi}^\sigma(x) + \tilde{\xi}^\sigma(x,t)
\end{array} \right.
\label{EqLangevin}
\end{align}
%Written in the lab frame,
%\begin{align}
%    \partial^2_t\psi = -\psi + \frac{4\epsilon}{3}|\psi|^2\psi - g\mathbf{\nabla}^2\psi -\delta\cos(\omega_D t)\psi -\eta\partial_t\psi + \xi
%    \label{EqLangevinLab}
%\end{align}
For simplicity, we limit ourselves to the case with particle-hole symmetry
\footnote{It is known that the Langevin equation of nonequilibrium superconductor can have a KPZ term which changes the universality class for dimensions lower than 3. For us, particle-hole symmetry forbids the KPZ term. Another interesting situation where the KPZ term needs modification is when the total charge of the system is strictly conserved.} %We study this case in future works.}
and focus on the limit $\eta\ll \delta$, $T_\text{eff}\rightarrow 0$. We simulate Eq.~\ref{EqLangevin} and the numerical results match well with our theoretical predictions (Fig.~\ref{FigOrbitSpectrum}-\ref{FigEMresponse}). See Appendix~\ref{AppendixNumerics} for details of the numerical simulation.

In three dimensions, where superconductors have true long range orders, we expect that the nonequilibrium PDSC phase is stable for $T_\text{eff} < O(\delta)$, which sets the energy scale for phase fluctuations but how this condition constrains the temperature of the bath and the coupling to the bath remains to be studied. 
%We shall discuss the nonequilibrium physics and the phase diagram of the parametrically driven Bose Hubbard model in detail in future works.

\section{Experimental signatures}

\begin{figure}[htb]
\begin{center}
\includegraphics[width=0.5\textwidth]{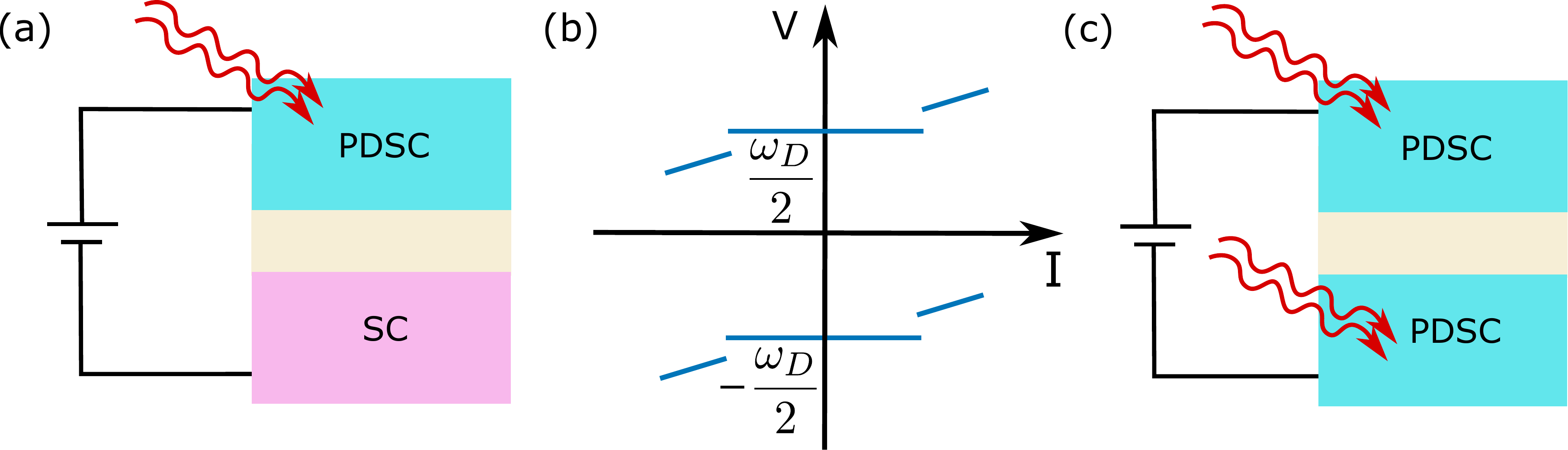}
\caption{(a) PDSC-SC Josephson junction. Because of the oscillation of the order parameter at frequency $\omega_D/2$ in PDSC, we expect a DC Josephson current when the bias voltage $V=\pm\omega_D/2$ (Eq.~\ref{EqPDSCSCJJ}). (b) Schematics of the DC I-V curve of the PDSC-SC Josephson junction in the vicinity of $V=\pm\omega_D/2$. The I-V relation depends on details away from $\pm\omega_D/2$. (c) PDSC-PDSC Josephson junction. We expect a Josephson current at $V=0,\pm\omega_D$ (Eq.~\ref{EqPDSCPDSCJJ}).}
\label{FigJosephson}
\end{center}
\end{figure}

We propose two Josephson-tunneling experiments to detect the PDSC phase. First, a Josephson junction (Fig.~\ref{FigJosephson}(a)) between a PDSC and an equilibrium superconductor with a constant order parameter $\psi_0 = |\psi_0|e^{i\phi_0}$. With a Josephson coupling $J_c\psi^*\psi_0 + h.c.$ and an applied voltage $V$, the supercurrent through the junction is
\begin{align}
I=&iJ_c\psi^*e^{iVt}\psi_0 - iJ_c\psi e^{-iVt}\psi_0^*\\
= &\sqrt{2J_0}J_c|\psi_0|\cos(\phi-\phi_0-(\omega_D/2+V)t) \nonumber\\
&- \sqrt{2J_0}J_c|\psi_0|\cos(\phi-\phi_0+(\omega_D/2-V)t).
\label{EqPDSCSCJJ}
\end{align}
For a conventional Josephson junction, DC Josephson current appears at zero bias voltage. For a PDSC-SC junction, DC Josephson current occurs exactly at $V=\pm\omega_D/2$. While $\omega_D$ can be in the infrared frequency, when $V\simeq \pm\omega_D/2$, there is a low-frequency current easy to detect electronically.

In some cases, the direct Josephson coupling does not exist. For example, in YBCO, it is proposed that the boson field itself is an effective description of pairing fluctuations at nonzero momentum~\cite{PhysRevB.101.064502}; therefore a direct coupling to another superconductor is forbidden by translation symmetry. To detect the PDSC phase in this situation, we propose making a Josephson junction between two copies of the material (Fig.~\ref{FigJosephson}(b)), driven by external fields with amplitude $\delta\cos(\omega_Dt)$ and $\delta\cos(\omega_Dt+2\theta_0)$ respectively.  The two resulting PDSC samples have amplitude $\psi_1=\sqrt{2J_0}\sin(\omega_Dt/2)e^{i\phi_1}$ and $\psi_2=\sqrt{2J_0}\sin(\omega_Dt/2+\theta_0)e^{i\phi_2}$ respectively. The Josephson current is
\begin{align}
I=&iJ_c\psi_1e^{iVt}\psi_2 + h.c.\\
=& 2J_0J_c\cos\theta_0\sin(\phi_1-\phi_2-Vt) \nonumber\\
&-2J_0J_c\cos(\omega_Dt+\theta_0)\sin(\phi_1-\phi_2-Vt)
\label{EqPDSCPDSCJJ}
\end{align}
which contains an ordinary AC-Josephson current and extra contributions at frequencies $V\pm\omega_D$.

\section{Discussion and Conclusion}

In this work, we propose a period-doubled superconducting steady state which emerges under periodic driving and  discuss experimental signatures which differentiate this intrinsically non-equilibrium superconductor from conventional ones.
%It is a challenging and rewarding task to detect this phase experimentally. 
In most existing experiments, the phonon oscillations, which serve as the periodic drive, last a few picoseconds~\cite{cavalleri2018photo}; at this timescale there is no evidence for a steady state. Theoretically, it has been predicted that the transient state exhibits a superconducting-like AC response but no Meissner effect~\cite{PhysRevB.104.054512}. For $\K3C60$, however, the pump laser itself was extended to a few hundred picoseconds and signatures of a quasi-steady state lasted for nanoseconds~\cite{budden2021evidence}; therefore, making it a promising case to test for PDSC. In materials close to the Mott transition, for example, $\ETBr$, our theory can also be applied to fractionalized bosonic chargons~\cite{PhysRevB.104.L241112}, where the physical conductivity depends on both the bosonic conductivity and the spinon conductivity.

Our theory can also be tested directly using cold atoms in optical lattices. Starting from a Mott insulator, our theory predicts that adiabatically turning on a periodic modulation of the height of the optical lattice at the frequency about twice of the Mott gap results in a time-dependent condensate. The number of bosons condensing at the zero momentum is proportional to $|\psi|^2=2J_0\sin^2(\omega_Dt/2)$. In order to realize this phase, there must not be a harmonic trap which pulls the excited particles to the center and push the holes outside, resulting in phase separation~\cite{navon2021quantum}. 

Finally, we comment on some questions our work raises regarding time crystals in open systems at $T>0$.  
Previous work has pointed out that TTSB is often unstable in systems which are in contact with a $T>0$ bath~\cite{bennett1990stability, yao2020classical}. This is because in non-equilibrium, a domain wall between two phases of the TTSB (here $\theta \to \theta + \pi$) can generically experience a linear force which causes them to nucleate and grow. 
However, in the present model a phase slip in the TTSB order $\theta \to \theta + \pi$ is homotopically equivalent to a phase slip $\phi \to \phi + \pi$ in the $U(1)$ order parameter. Because  $U(1)$ fluctuations experience only a gradient penalty, it will thus be energetically favorable for the former to convert to the latter.
The $U(1)$ symmetry then forbids a linear force on the domain wall, and we may expect TTSB order is preserved for temperatures below the 3D XY transition $T_c$. 
Thus, we conjecture that the 3D model exhibits infinitely long-lived TTSB for $T < T_c$.

This $U(1)$-protected TTSB differs in several respects from other scenarios in which oscillations ``piggyback'' off a conserved $U(1)$ quantum number~\cite{else2020discrete}, as for example the phase of a condensate at finite chemical potential, $H = - \mu N, [\phi, N] = i$. In the ``piggyback'' scenario, the trajectory of the oscillations (e.g. $ \phi(t) = -\mu t + \phi_0$) is simply an orbit under the $U(1)$ symmetry; in the present model the oscillation in $\theta(t)$ is not.  
This presumably affects the stability of the model when the symmetry is explicitly broken by a term like $\cos(2 \phi)$. In the ``piggyback'' scenario, oscillations will be rapidly destroyed~\cite{else2020discrete}.
In the present case the TTSB should remain activated, i.e. persist out to an exponentially long time in the inverse temperature, due to the effective energy penalty for both $\theta$ and $\phi$ phase slips. 
It would be interesting to explore these conjectures using  finite temperature Langevin simulations.

\section{Acknowledgment}
We thank Patrick A. Lee, Ehud Altman, Yantao Wu, Francisco Machado, Bingtian Ye and Stefan Divic for helpful discussions. This work was supported in part by the U.S. Department of Energy, Office of Science, National Quantum Information Science Research Centers, Quantum Systems Accelerator (QSA), the Gordon and Betty Moore Foundation (Grant GBMF8688), and the U.S. Department of Energy, Office of Basic Energy Sciences, Division of Materials Sciences and Engineering under Award No. DE-SC0019241.

\bibliographystyle{apsrev4-1}
\bibliography{ref.bib}

\appendix

\section{Numeric simulation of the electromagnetic response}
\label{AppendixNumerics}

In this appendix, we provide additional information about the numerical simulation.

\subsection{Discretization of the Langevin equation}

We first rewrite Eq.~\ref{EqLangevin} in the lab frame. 
\begin{align}
&\left\{ \begin{array}{l}
\partial_t\psi^\sigma(x) = \frac{\delta H_\text{con}}{\delta \pi^{\sigma}(x)}\\ \partial_t\pi^{\sigma}(x) = -\frac{\delta H_\text{con}}{\delta \psi^{\sigma}(x)}-\eta\pi^\sigma(x) + \xi^\sigma(x,t)
\end{array} \right.
\label{EqLangevinlab}
\end{align}
where $\xi$ satisfies $\<\xi^\sigma(x,t)\xi^{\sigma'}(x',t')\>=2\eta T_\text{eff}\delta^{\sigma,\sigma'}\delta(x-x')\delta(t-t')$ in the long wavelength limit.

We simulate this Langevin equation on a discrete lattice with discrete time steps. To do this, we rewrite the continuum Hamiltonian (Eq.~\ref{EqHamiltonian}, Eq.~\ref{EqHamiltonianApsi}) as

\begin{align}
H_\text{dis} = \sum_{\mathbf{r}}&\frac{1}{2}|\pi_\mathbf{r}|^2  +V(\psi_\mathbf{r}) +\frac{\delta}{2}\cos(\omega_D t)|\psi_\mathbf{r}|^2\nonumber\\
&+ \frac{g}{2}|\psi_\mathbf{r}-e^{iA_{\mathbf{r}+\hat{x}/2}}\psi_{\mathbf{r}+\hat{x}}|^2 + \frac{g}{2}|\psi_\mathbf{r}-e^{iA_{\mathbf{r}+\hat{y}/2}}\psi_{\mathbf{r}+\hat{y}}|^2,
\label{EqDiscreteHamiltonian}
\end{align}
where $\mathbf{r}$ takes value from a $L\times L$ square lattice, with periodic boundary condition, $\psi_\mathbf{r}=\psi^a_\mathbf{r}+i\psi^b_\mathbf{r}$ is the complex boson filed, $\pi_{\mathbf{r}}=\pi^a_\mathbf{r}+i\pi^b_\mathbf{r}$ is the corresponding momentum. $A_{\mathbf{r}+\hat{x}/2}=\int_\mathbf{r}^{\mathbf{r}+\hat{x}}\vec{A}\cdot d\mathbf{r}'$ is the discrete version of the x component of the vector potential, similarly $A_{\mathbf{r}+\hat{y}/2}=\int_\mathbf{r}^{\mathbf{r}+\hat{y}}\vec{A}\cdot d\mathbf{r}'$. The Langevin equation for each discrete time step is

\begin{align}
&\left\{ \begin{array}{l}
\Delta\psi^\sigma_\mathbf{r} = \frac{\delta H_\text{dis}}{\delta \pi^{\sigma}_\mathbf{r}}\Delta t\\ \Delta\pi^{\sigma}_\mathbf{r} = -\frac{\delta H_\text{dis}}{\delta \psi^{\sigma}_\mathbf{r}}\Delta t-\eta\pi^\sigma_\mathbf{r}\Delta t + \xi^\sigma_{\mathbf{r},t,\Delta t}
\end{array} \right.
\end{align}
where $\Delta t$ is the discrete time step. We fix $\Delta t = 2\pi/80\omega_D$ for every simulation in this work. $\xi^\sigma_{\mathbf{r},t,\Delta t}\equiv\int_t^{t+\Delta t}\xi^\sigma_\mathbf{r}(t)dt$, where $\xi^\sigma_\mathbf{r}(t)$ is the random noise which satisfies $\<\xi^\sigma_\mathbf{r}(t)\xi^{\sigma'}_\mathbf{r}(t)\>=2\eta T_\text{eff}\delta^{\sigma,\sigma'}\delta_{\mathbf{r},\mathbf{r}'}\delta(t-t')$. To reproduce the correct expectation value for $\<\xi^\sigma_\mathbf{r}(t)\xi^{\sigma'}_\mathbf{r}(t)\>$, we set $\xi^\sigma_\mathbf{r}(t)$ as independent random numbers evenly distributed in $[-\sqrt{6\eta T_\text{eff}\Delta t},\sqrt{6\eta T_\text{eff}\Delta t}]$. We focus on the limit $T_{\text{eff}}\ll \delta$ in this work.

\subsection{Choice of the vector potential}

To simulate the response to an AC electric field, we set the vector potential to be

\begin{align}
&\left\{ \begin{array}{l}
A_{\mathbf{r}+\hat{x}/2}=A_0f(t)\sum_i\cos(\omega_it)\\ 
A_{\mathbf{r}+\hat{y}/2}=0
\end{array} \right.
\end{align}
We set $A_0=0.1$, $f(t)=\exp(-(2500\pi)^2/(\omega_D t)^2)$, and $\omega_i=\omega_D/500,3\omega_D/500,5\omega_D/500,7\omega_D/500,9\omega_D/500$. We measure the current density on each bond in the x direction in the time period $[20000T_D,30000T_D]$

\begin{align}
    j_{\mathbf{r}+\hat{x}/2}(t)
    = &g\epsilon_{\sigma\sigma'}\psi^\sigma_{\mathbf{r}+\hat{x}}(t)\psi^{\sigma'}_{\mathbf{r}}(t)\nonumber\\
    &- g\delta_{\sigma,\sigma'}\psi^\sigma_{\mathbf{r}+\hat{x}}(t)\psi^{\sigma'}_{\mathbf{r}}(t)A_{\mathbf{r}+\hat{x}/2}(t)
\end{align}
and calculate the conductivity at each frequency $\omega_i$.

To simulate the steady state with a constant magnetic field, we take a $L\times L$ periodic lattice, $\mathbf{r}=(x,y), x,y=0,\dots L-1$, and set the vector potential as follows

\begin{align}
&\left\{ \begin{array}{l}
A_{\mathbf{r}+\hat{x}/2}=2\pi y N_\text{flux}/L^2\\ 
A_{\mathbf{r}+\hat{y}/2}=0,\  y\neq L-1\\
A_{\mathbf{r}+\hat{y}/2}=-2\pi x N_\text{flux}/L,\  y= L-1
\end{array} \right.
\end{align}
such that there is $2\pi N_\text{flux}/L^2$ magnetic flux through each plaquette, and $N_\text{flux}$ total flux on the torus.

\vspace{5mm}

\section{Driven bosons coupled to a bath of acoustic phonons}
\label{AppendixPhononBath}

In this appendix, we discuss a specific coupling between the periodically driven bosons and a heat bath consisting acoustic phonons. We show that under this coupling the steady state has an effective temperature close to the temperature of the bath.

We write the boson field and its canonical momentum in terms of the particle annihilation operator $a_k$ and the hole creation operator $b_{-k}^\dagger$

\begin{align}
\psi_k &\equiv \frac{1}{\sqrt{E_k}}(a_k + b_{-k}^{\dagger})\\
\pi_k &\equiv \sqrt{E_k}(-i a_k + ib_{-k}^{\dagger}). &
\end{align}
Consider an acoustic phonon field $\chi$, and the following interaction 
\begin{align}
    H_I=\sum_{k,q}f(q)\chi_q (a^\dagger_{k+q}a_k+b^\dagger_{k+q}b_k)+h.c.
\end{align}
In order to find the steady state of the driven bosons, we would like to rewrite the interaction in the rotating frame, in terms of $\tilde{a}_k$ and $\tilde{b}_k$ defined as follows

\begin{align}
\tilde{\psi}_k &\equiv \frac{1}{\sqrt{E_k}}(\tilde{a}_k + \tilde{b}_{-k}^{\dagger})\\
\tilde{\pi}_k &\equiv \sqrt{E_k}(-i \tilde{a}_k + i\tilde{b}_{-k}^{\dagger}). &
\end{align}

For bosons near the band minimum, since we have set the band gap to 1,

\begin{align}
    a_k = (\sqrt{E_k}\psi_k + i\pi_k/\sqrt{E_k})/2\simeq (\psi_k + i\pi_k)/2\nonumber\\
    =e^{-i\omega_D t/2}(\tilde{\psi}_k+ i\tilde{\pi}_k)/2\simeq e^{-i\omega_D t/2}\tilde{a}_k
\end{align}
Similarly, $b_k \simeq e^{-i\omega_D t/2}\tilde{b}_k$. Thus

\begin{align}
    \tilde{H}_I(\chi,\tilde{a},\tilde{b},t) &= H_I(\chi,a,b)\nonumber\\
    &\simeq \sum_{k,q}f(q)\chi_q (\tilde{a}^\dagger_{k+q}\tilde{a}_k+\tilde{b}^\dagger_{k+q}\tilde{b}_k)+h.c.,
\end{align}
which has no explicit time dependence even in the rotating frame. Thus, the ``total energy" measured in the rotating frame is approximately conserved and the steady state has an effective temperature close to the temperature of the bath. Physically, when the acoustic phonon has a low temperature, it helps the excited bosons relax to zero momentum, hence reaching the minimum of the effective Hamiltonian.

\end{document}